%% file: main.tex
\documentclass{article}
\input{packages}

\title{NuGraph2 with Explainability: Post-hoc Explanations for Geometric Neural Network Predictions}
\date{\today}
\input{authors}

\begin{document}

\maketitle

\begin{abstract}

    With the growing popularity of artificial intelligence used for scientific applications, the ability of attribute a result to a reasoning process from the network is in high demand for robust scientific generalizations to hold.
    In this work we aim to motivate the need for and demonstrate the use of post-hoc explainability methods when applied to AI methods used in scientific applications.
    To this end, we introduce explainability add-ons to the existing graph neural network (GNN) for neutrino tagging, NuGraph2. 
    The explanations take the form of a suite of techniques examining the output of the network (node classifications) and the edge connections between them, and probing of the latent space using novel general-purpose tools applied to this network.
    We show how none of these methods are singularly sufficient to show network "understanding", but together can give insights into the processes used in classification. While these methods are tested on the NuGraph2 application, they can be applied to a broad range of networks, not limited to GNNs. The code for this work is publicly available on GitHub at \url{https://github.com/voetberg/XNuGraph}.
\end{abstract}


\section{Introduction}

As the usage for artificial intelligence (AI) becomes more pervasive across all domains of science and society, the need for explainability of AI methods, used e.g. to make a scientific discovery or to drive certain decisions in medical procedures, becomes more urgent than ever. Such methods are often regarded as black-boxes, whose power remains obscure even to the understanding of experts. The lack of explainability, thus, makes it, on the one hand, difficult to communicate and accept their outcome, and on the other hand, prevents experts from learning which features in their data played a significant role. This can leave experts unaware of potential unknown  correlations or, worse, of unexpected shifts in the data distributions.

Past work in the domain of AI explainability has brought significant advancements for convolutional neural networks (CNNs), with the development of tools and techniques that can be broadly applied to different applications (see Sec.~\ref{sec:relatedwork}). In the past years, Graph Neural Networks (GNNs) have emerged as a powerful generalization of CNNs that operate on data layouts that are not necessarily image-like but are rather described as a graph. This graph data structure consists of a set of nodes and a set of edges connecting them. Explainability tools for GNNs are less widely adopted than those for CNNs and, due to the larger flexibility of graph-based data structures, they often are less generalizable.
In this paper, we explore explainability approaches in the context of a specific GNN for High Energy Physics (HEP) experiments, aiming at providing insights, ideas, and methods that can be valuable for the larger GNN context. 
There exist a general set of questions one can ask of neural network explainability methods, and we narrow these as incremental steps towards a wider understanding of GNNs.
We are interested in probing the network behavior in terms of information flow and learned notions.
In this paper we address a set of specific questions, and test or develop tools for extracting this information from the NuGraph GNN~\cite{nugraph2}, such as: "Why did the network get a prediction wrong or correct?",  "What information is captured in the network's internal representation, and how does the training process influence it?", "How does the information flow across the graph edges, and what is the role of nodes with different edge degree?", "What is the role of network hyperparameters, such as the number of message-passing iterations?"
Ultimately, we aim to address more fundamental questions, including: "Does it correctly encode physical laws present in the training data without being explicitly told?", "How does the network ``conceptualize`` different features of inputs?", "How does the network utilize positional information versus edge information?", "How does the network, if at all, learn derived graph-wide features?". 
The work presented here represents a fundamental step in the direction towards such ultimate questions.
We present multiple techniques that aim to address overlapping interpretations of these questions.

We also make a distinction between ``interpretable`` and ``explainable`` results. 
While these terms are often interchangeable and their definitions fluid, it is generally considered that these terms do not describe exactly analogous ideas. 
For the sake of this paper, we adopt the definition of interpretable put forward by Doshi-Valez and Kim \cite{doshivelez2017rigorousscienceinterpretablemachine} - "ability to explain or to present in understandable terms to a human". 
While a large part of interpretable and explainability works aim to make a `white-box` network as a proxy \cite{lipton_mythos}, we follow the narrower definition of "explainability" give by Gunning and Aha \cite{gunning_aha_explainable}, where explainability is a set of methods used \textit{after} the network is trained. 
We aim to give attributions for prediction results from the network, not to create a transparent (interpretable) network.
Our methods are all used \textit{post-hoc}, and examine the de-facto way the network has trained. 

In what follows, we first present our specific application, NuGraph (Sec.~\ref{sec:nugraph}). Then we survey and summarize the literature  in the domain of AI explainability (Sec.~\ref{sec:relatedwork}), and present our attempts at adopting  GNN explainability methods in our application domain (Sec.~\ref{sec:netexpl}). Finally, we present a few methods we developed to gain further insights into our application (Sec.~\ref{sec:rewiring}, Sec.~\ref{sec:latentspace}, Sec.~\ref{sec:probes}). We conclude summarizing our work in Sec.~\ref{sec:conclusions}.

\section{Related Work}
\subsection{Case Study Application - NuGraph}
\label{sec:nugraph}

NuGraph2~\cite{nugraph2} is a GNN for the classification of energy deposits from liquid argon time projection chamber (LArTPC) neutrino experiments. LArTPC detectors record the trajectories of charged particles, produced by the interaction of neutrinos with the argon or by cosmic ray backgrounds. These particle ionize the argon along their path and, due to the uniform electric field applied to the detector, the ionization electrons drift towards the anode plane. Here they are detected in the form of electric pulses on sense wires arranged in different planes, typically three. Each plane records a two-dimensional trajectory in wire space and time, and due to the different direction of the wires, when these are combined across planes the full three-dimensional trajectory can be recovered. After the electric signals on the wires are pre-processed, pulses are fit as one or more Gaussian distribution, where each distribution represents the energy deposition of a single particle, or "hit". Hits contain information such as the wire and time of detection, the integral of the Gaussian (proportional to the energy that was deposited by the particle), and its width.

Hits represent the nodes of the graph, and their properties are the input features to the GNN. For every recorded neutrino interaction ("event"), in each wire plane a graph is constructed by connecting hits using Delaunay triangulation based on their wire and time coordinates. Hits in different planes are also connected using the SpacePointSolver \cite{abi2020deepundergroundneutrinoexperiment} algorithm, which determines 3D space points based on the intersection of wires and the time coincidence of hits. These space points are used in the graph as "nexus" nodes, bridging the information across planes and allowing the GNN to be exposed to 3D information. The GNN classifies the hits in two ways: a "filter" decoder classifies hits as either from particles produced by a neutrino interaction (signal) or from noise or cosmic rays (background); a "semantic" decoder classifies hits according to the type of particle they originated from. Semantic categories are minimum ionizing particle (MIP), highly ionizing particle (HIP), shower, Michel electron, and diffuse. The GNN will be able to distinguish between these classes based on the graph topology and the hit input features. For instance, both MIP and HIP particles lead to track-like topologies (i.e., hits aligned along lines) but they differ in the amount of energy deposited, which is captured in the integral of their hits. Shower particles, instead, show a different topology, with a cascade developing in branches with a tree-like structure.

NuGraph2 is trained on the MicroBooNE public data sets~\cite{abratenko_2023_8370883,abratenko_2022_7261921}, which feature simulated neutrino interactions from the Booster Neutrino Beam overlaid on top of off-beam data collected with the MicroBooNE detector~\cite{MicroBooNE:2016pwy}. Results presented in this paper are also evaluated on these data sets. A more complete description of these data sets can be found in~\cite{Cerati:2023rtv}.
NuGraph2 achieves an overall accuracy of 98\% and 95\% for the filter and semantic decoders, respectively. The performance of the semantic decoder is not uniform across the different categories, so that the decoder performs better on the most represented category in the training dataset (MIP) and worst on the least represented category, Michel. Usually, mistakes take place when the network is unsure about the classification of a group of hits, resulting in a mixed classification of that group across multiple categories. In rare circumstances, large scale failures can also happen, where most of the hits in an event (or in a particle in the event) are incorrectly classified. Understanding the origin of these mistakes is a fundamental step that can drive developments aiming at further improvements of the model.

For this reason explainability is a priority for our work, and the GNN and AI/ML communities at large. It is however worth pointing out the specific features that characterize NuGraph compared to a "typical" GNN in the literature. 
\begin{itemize}
    \item As each neutrino interaction leads to a unique set of hits, the graph nodes and edges are not fixed.
    \item Delaunay triangulation leads to a graph that is connected (i.e. a path exists that connect all of pairs of nodes), but where edges don't necessarily have physical interpretation, and where some "hub" nodes can have a large number of edges. We rely on the attention mechanism to dynamically identify meaningful edges.
    \item NuGraph2 is a heterogeneous network, as it features two kind of nodes (planar and nexus) with different properties.
    \item NuGraph2 is a multi-category and multi-decoder network, with a latent space capturing information relevant for different tasks.
\end{itemize}
These unique (or unusual) properties imply that some of the standard methods for GNN explainability may not be directly applicable to NuGraph2, and that our work will extend the functionalities of (or in general the questions asked to) GNN interpretability tools.

\subsection{Literature Review}
\label{sec:relatedwork}

The study of explainablity in graph neural networks is not a novel field. 
Taxonomy studies \cite{yuan_explainability_2022} \cite{dai_comprehensive_2023} breaks down the majority of GNN-focused methods into two broad categories for the analysis for trained networks - explanations for single inference result and explanations for the model itself. 
Some notable methods include `GNNExplainer` \cite{ying_gnnexplainer_2019} and `PGExplainer` \cite{luo2020parameterizedexplainergraphneural}
which use masking methods to determine which parts of the graph most influence a node's inference results, `SubgraphX` \cite{yuan2021explainabilitygraphneuralnetworks}, which uses a tree search method with Shapley evaluation metrics, and `GraphLIME` \cite{huang2020graphlimelocalinterpretablemodel} which adapts the popular method `LIME` \cite{ribeiro2016whyitrustyou} to GNNs. 
All of these are focused on inference-level explanations. 
The best known whole-model explanation method is the generative `XGNN` \cite{yuan_xgnn_2020} method, which endeavors to encode the "rules" of the network to produce a small class of subgraphs representing the model's "understanding" of the input data.

To build off these methods, and to adapt this method to our unique constructed heterogeneous graph structure, we draw from the idea of TCAV \cite{kim2018interpretabilityfeatureattributionquantitative} to assign groups of behaviors and Linear Probes \cite{alain_understanding_2018} to access the network's understanding at different stages of inference. 
This is critically important to a message passing network with an empirically determined number of message passing steps such as NuGraph2.

\section{Methods and Results}

\subsection{Network Explanations}
\label{sec:netexpl}
 Much of the existing literature for post-training graph network explainability focuses on geometric situations with physical edges, or edges with some meaning relative to the nodes. 
By contrast, NuGraph2's edges are all constructed algorithmically using Delaunay triangulation. 
The process of reconstructing the tracks to align with physical constraints would require knowledge of the identity of particles leaving the tracks, and render the identification network redundant.
This caused us to explore applications of existing methods that can work around the nonphysical edges, such as GNNExplainer \cite{ying_gnnexplainer_2019}.
 
GNNExplainer works by optimizing masks (through gradient descent) applied on the network to both a) include the most relevant elements of a graph through an entropy measure, and b) minimize the number elements included in the final graph. 

We make two adjustments to the existing GNNExplainer algorithm: 
\begin{enumerate}
    \item We extend the existing implementation supplied by PyTorch-Geometric \cite{FeyLenssen2019} to produce multiple masks optimized jointly across different planes of our heterogeneous graphs. 
    \item We added the capability of working with heterogeneous graphs, i.e. planar and nexus nodes in NuGraph2.
    \item We use inference results and ground truth information to produce subgraphs highlighting specific failure modes of NuGraph2.
\end{enumerate}


A total of 6 masks (one for each planar graph, and one for each set of connections between the planar graph and the nexus layer) are initialized and jointly optimized over graphs $G$ through the entropy minimization equation $\mathcal{L} = H(Y)-H(Y|G = G_S)$. 
In our case, $G_S$ is defined as the subgraph with all 6 masks applied.  
Afterwards, we then examine the individual subgraphs constructed using the ground truth of our data. 
Through this we aim to examine two classes of questions. 
\begin{enumerate}
    \item How does the network's understanding of individual classes rely on its connections to other classes within the graph? 
    \item How do correct and incorrect classifications arise within the same graph?
\end{enumerate}

Both of these questions are addressed by limiting the entropy loss to only target specific nodes within a given graph, and  both are done training the explainer algorithm separately for each sub-task. 
The first recasts $H(Y)$ to be $H(f(X|Y_\texttt{true} = c)), c\in [\texttt{HIP}, \texttt{MIP}, \texttt{Shower}, \texttt{Michel}, \texttt{Diffuse}]$ (NuGraph2's semantic classes); where $f$ is our network.
The second applies a similar constraint by requires either $H(Y_\texttt{pred}=Y_\texttt{true})$ or $H(Y_\texttt{pred}\neq Y_\texttt{true})$. 

In order to judge the effectiveness of our different variants of GNNExplainer, we use two main metrics, Fidelity$_{+}$ and Fidelity$_{-}$, with values ranging between 0 and 1 \cite{yuan_explainability_2022}.
These metrics can be used to examine how much the inference of a graph is impacted after the explanation method is applied. Fidelity$_{-}$ indicates that removing unimportant information either does not impact the result at all, or improves the result.
It can be formalized as: 
Fidelity$_{-} = \frac{1}{N} \sum^{i=1}_{N} (f(G_i) - f(G^{m_i}_{i}))$ over a set of graph samples of size $N$ with a masks $m \in M$, thus implying that a lower value denotes a more useful explanation. 
Conversely, Fidelity$_{+}$ removes the nodes and/or edges included in the explanation from the graph, through the formalism Fidelity$_{+} = \frac{1}{N} \sum^{i=1}_{N} (f(G_i) - f(G^{1-m_i}_{i}))$. 
This makes the inverse of Fidelity$_{-}$'s assertion true, where a value closer to 1 denotes an explanation that encapsulates all information important to the network's result.  
We take a subset of our testing data for this analysis, trying to identify "interesting" examples of network failures. 
To filter our results, we take the 8 poorest performing graphs on two different criterion (16 total). 
This smaller subset was selected to allow for visual inspection of each explanation, and as GNNExplainer optimizes each mask individually, a large dataset is not required to see results. 
Criterion one identifies the graphs with the largest number of nodes with true labels "MIP" or "HIP" (the "track-like" classes) that were incorrectly labeled as "HIP" or "MIP", respectively. 
Criterion two first takes graphs with true "Michel" labels (the most underrepresented and poorest performing class), and then orders them by the largest number of either false positive or false negative Michel classifications. 

\begin{figure}
    \centering
    \includegraphics[width=0.7\linewidth]{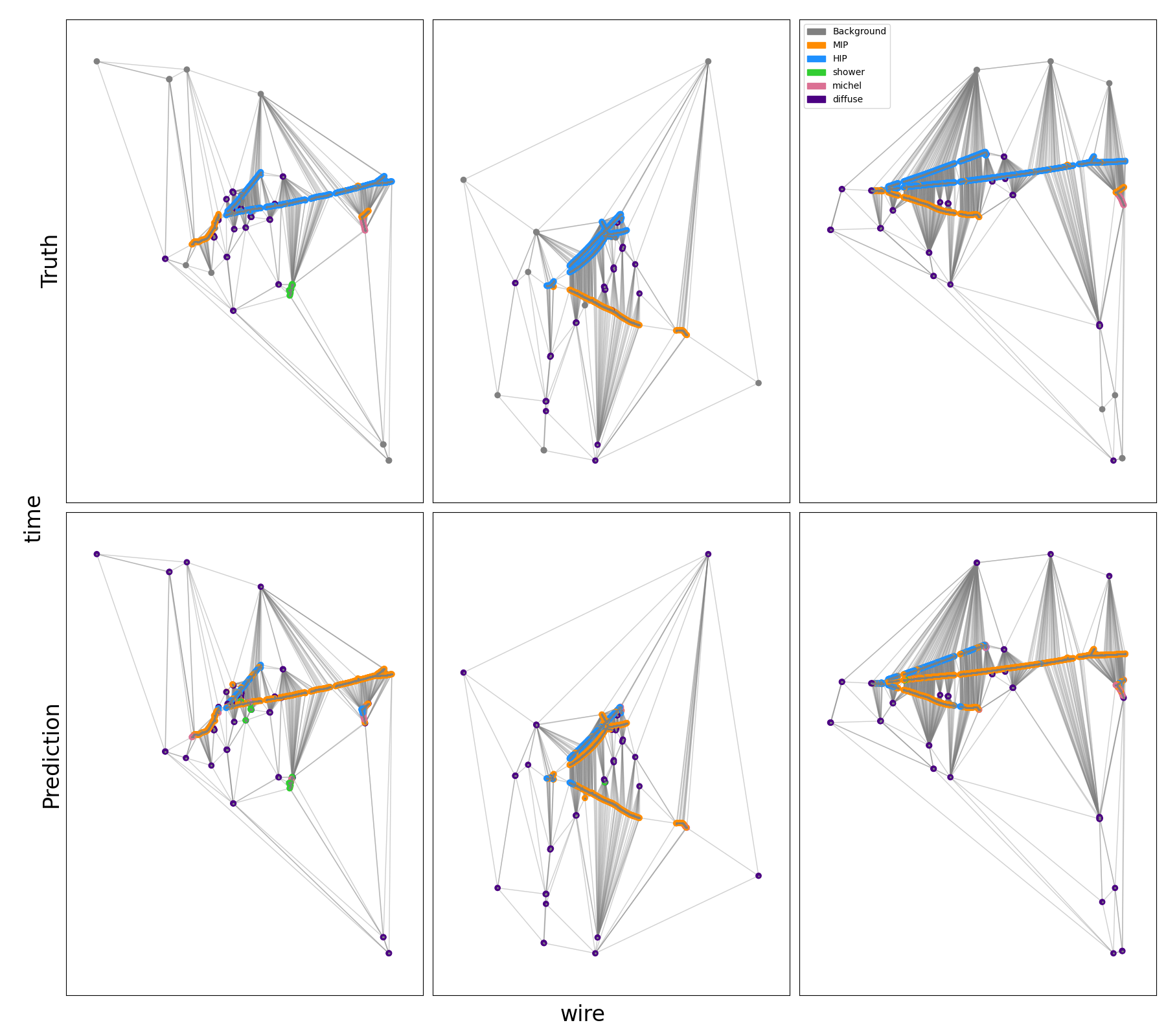}
    \caption{An example event used to investigate different GNNExplainer variants. The event shows a clearly misidentified track event typical of one of the common modes of failure in NuGraph2, where a central particle track in a group of three tracks is attributed incorrectly throughout the whole track. Each horizontal panel represents an event view from the preceptive of a different wire plane (ordered 'u', 'v', 'y')}
    \label{fig:example-event}
\end{figure}

\begin{figure}[h]
    \centering
    \includegraphics[width=0.8\linewidth]{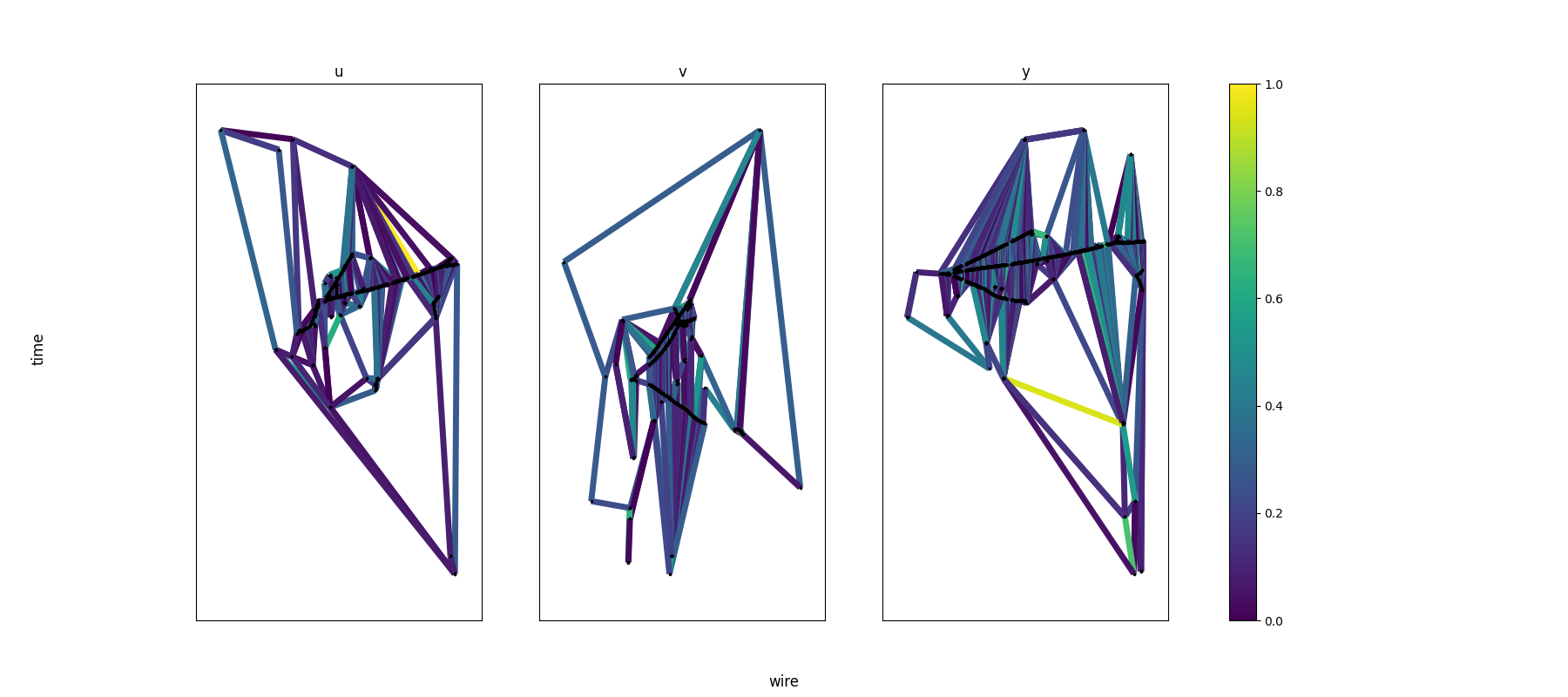}
    \caption{
    An example result of the GNN Explainer algorithm as applied to NuGraph input data. This specific graph is targeting the incorrectly classified hits as shown in figure \ref{fig:example-event} and attempting to identify the most important edges to causing the incorrect classification of the central track. The figure above ranks the edges based on their importance to predicting node classifications, color coded from least important (most blue) to most important (most yellow). The general uniformity of the importance ranks renders this method inoperable for the purposes of NuGraph2 input data. This is supported by the data in table \ref{tab:gnnexplainer-metrics}.}
    \label{fig:event_explaination}
\end{figure}

The above approaches have two main limitations. 
The first is that the gradient based optimization method central to the discovery of important subgraphs in GNNExplainer cannot reach satisfactory minima in NuGraph2's graphs.
This is shown in table \ref{tab:gnnexplainer-metrics}, where none of the methods reach a sufficient Fidelity$_\pm$ score to declare the subgraphs representative of the graph. 
In fact, having both Fidelity$_{+}$ and Fidelity$_{-}$ values close to one indicates that all graph features are considered important for the observed result. Since the method is not able to identify which aspects are less relevant, explanations provided lack in conciseness and are ultimately not providing useful insight related to the specific outcome (Fig.~\ref{fig:event_explaination}).
We hypothesize that methods that define masks describing relevant sub-graphs for the observed outcome struggle with NuGraph2 due to the large degree of redundancy in the edge connections. We verify this hypothesis in Section \ref{sec:rewiring}.

The second limitation has to do with the combination of the base method and the information we want to extract. GNNExplainer strives to make subgraphs that explain small groups of nodes or edges in the graph, and while it can be applied to whole graphs, it cannot capture a comprehensive explanation of the whole network's "understanding" or encoding of the problem as a whole. 
GNNExplainer, in the case of NuGraph2, is limited to semi-local explanations of singular graphs, and as a consequence relies on the practitioner to find examples of certain kind of behaviors seen in the data, with no guarantee the inferences made from a singular graph extend to the problem as a whole. 
To address this limitation, in Sections~\ref{sec:latentspace} and \ref{sec:probes} we investigate how the network is able to distinguish between different general concepts.

In summary, while GNNExplainer and other node and edge attribution methods are generally useful, especially for identifying the cause of misclassifications, they do not extend to the particulars of NuGraph, nor all graph networks. 

\begin{table}[h]
    \centering
    \begin{tabular}{ c c | c | c }
        &  & Fidelity$_{-}$ & Fidelity$_{+}$\\
        \hline\hline
        Base GNN Explainer &  & 0.960 &  0.962 \\
        \hline
        Class Based Subgraphs & HIP &  0.951 & 0.948 \\
            & MIP & 0.951 & 0.948 \\
            & Shower &  0.952 &  0.949 \\
            & Michel &  0.949 &  0.952 \\
            & Diffuse & 0.948 & 0.955 \\
        \hline
        Classification Result Subgraphs & Correct & 0.954 & 0.962\\
            & Incorrect & 0.960 & 0.955 \\

    \end{tabular}
    \caption{Table showing the results of the two main metrics for the different GNN explainer methods used on the event shown in figure \ref{fig:example-event} targeting edge explanations. Metrics are averaged over the 3 planes of the graph. }
    \label{tab:gnnexplainer-metrics}
\end{table}

\subsection{Graph Rewiring}
\label{sec:rewiring}

To further address the question of "How many edges are enough?" and investigate the inductive biases added by the Delaunay triangulation algorithm, we conduct graph rewiring and edge pruning. Delaunay triangulation is used to create a graph from the point cloud data. This wiring algorithm subdivides the convex hull of a set of points on a plane into triangles whose circumcircles (i.e. circles that pass through all the three vertices forming the triangle) don't contain any of the points. While connecting point clouds using Delaunay triangulation does not immediately lead to physically interpretable connections, it nevertheless carries learning biases due to the algorithm definition, such as:
\begin{itemize}
    \item \textbf{Locality}. Nearest-neighbors are by construction a preferred connection path with Delaunay triangulation. Long-range dependencies are present but less frequently created. In NuGraph2, isolated nodes act as hubs of long-range connections. This implies that information across distant nodes needs to travel through several message passing iterations or through hub connections.
    \item \textbf{Density-dependent node degree}. After the triangulation, there are typically sparse regions with low node degree, and densely connected regions with high node degree. Thus, nodes in sparse regions may suffer from under-smoothing (little information is aggregated about its neighborhood or context) while nodes in dense regions can suffer from over-smoothing (a large amount of  information from the neighborhood  is condensed, causing a loss of ``resolution``).
\end{itemize}

We evaluated the performance of the network under different edge pruning schemes and rewiring techniques listed below: 
\begin{itemize}
    \item Complete, randomized rewiring of the graph while conserving the total number of edges (e.g., $(i,j) \mapsto (k,l)$). This approach retains no information from the original wiring besides the total number of edges.
    \item Randomly permutation of the target nodes in the edge index tensor (e.g., $(i,j) \mapsto (i,k)$). This approach retains more information from the original wiring compared to the first approach. 
    \item Dropping a random edge with probability $p$.
    \item Limiting the maximum degree of each node to $M$ and randomly pruning excess edges.
    \item Uniformly sampling $M$ edges of nodes with degree greater than $M$ according to their length in the $wire \ vs \ time$ time space.
\end{itemize}
All the rewiring and pruning methods listed above were carried out in such a way that ensures that no nodes have a self-connection or are isolated. Out of all tried methods, only the  last approaches, which limited the maximum degree of each node, produced results as good as the baseline Delaunay wiring and thus are discussed more in-depth. 

\begin{figure}
    \centering
    \includegraphics[scale=0.34]{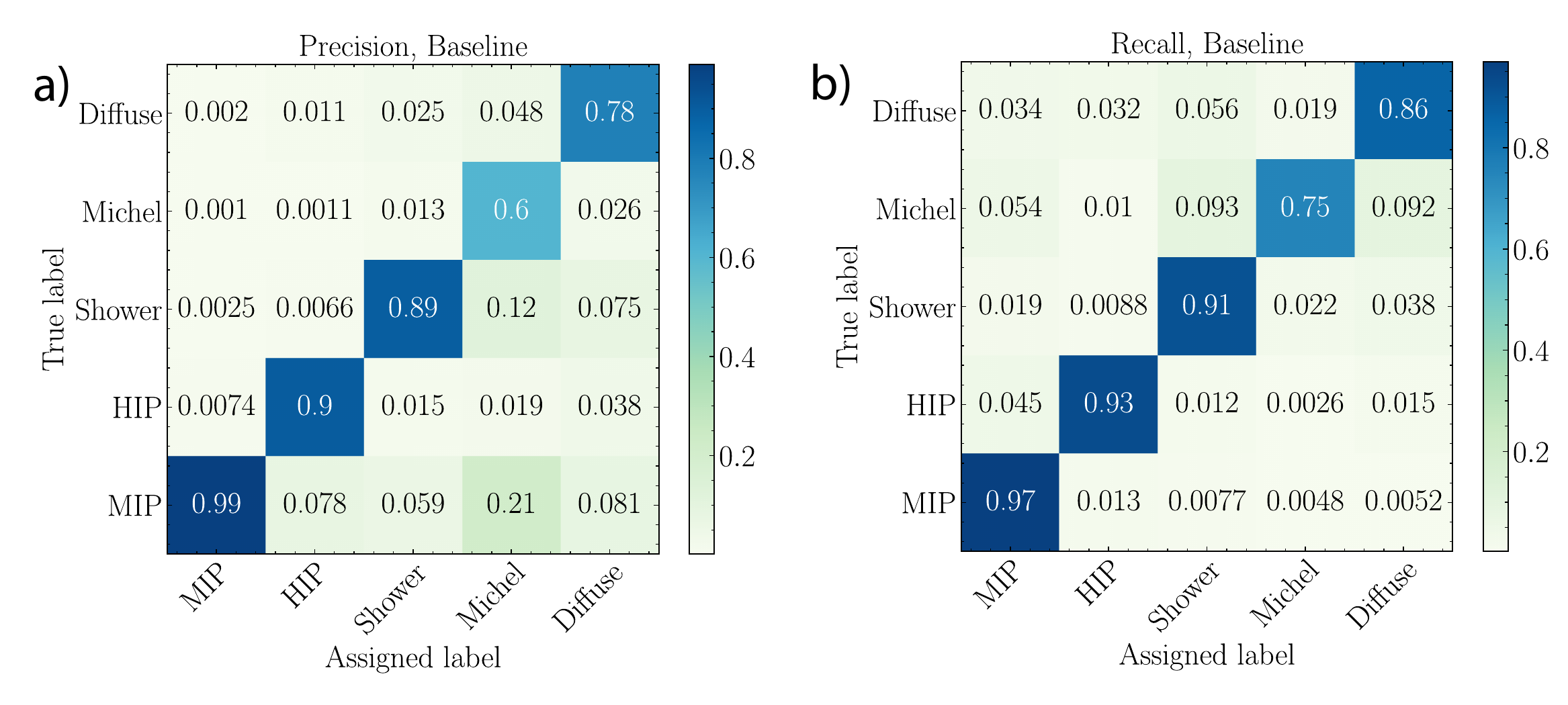}
    \caption{\textbf{a)} Precision and \textbf{b)} Recall confusion matrices for the baseline network.} 
    \label{fig:baseline}
\end{figure}

\begin{figure}
    \centering
    \includegraphics[scale=0.34]{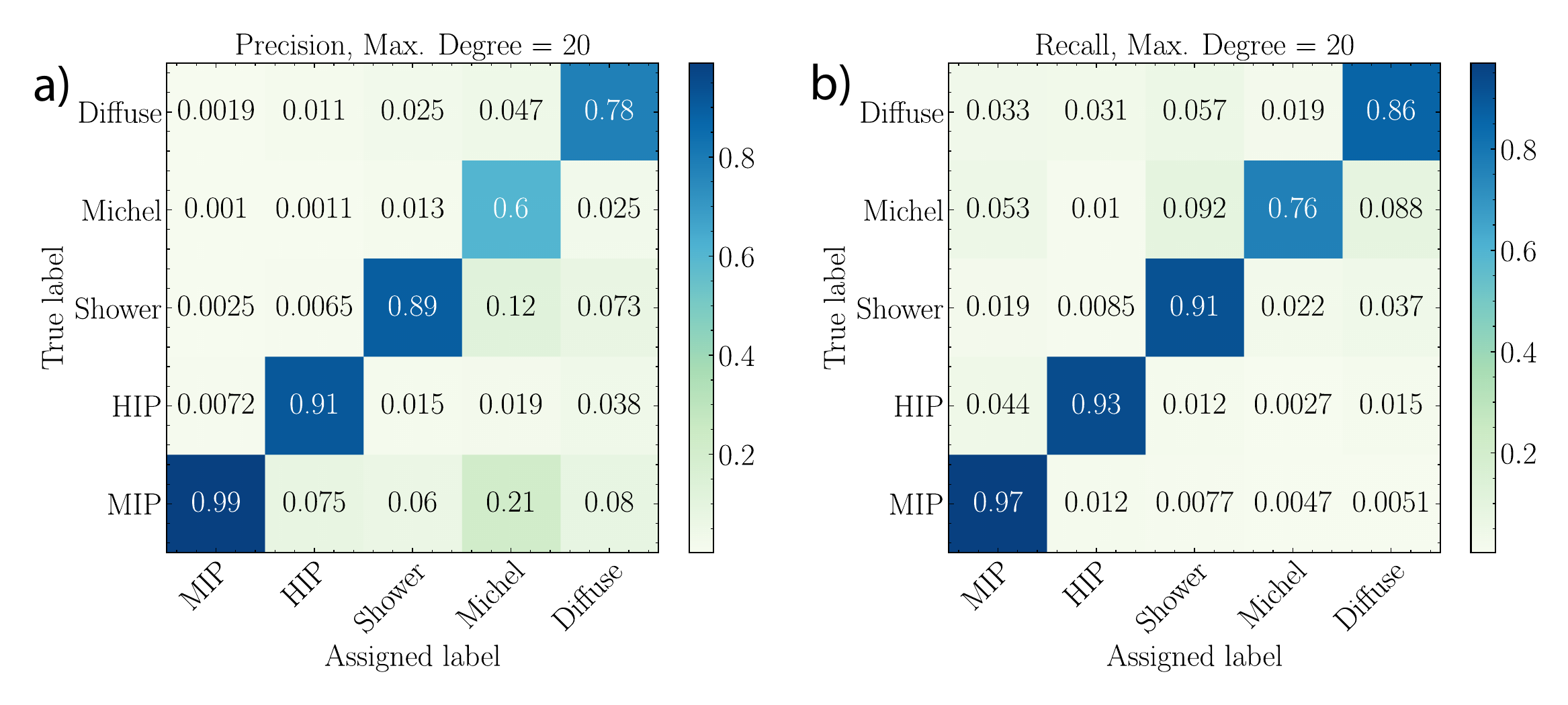}
    \caption{\textbf{a)} Precision and \textbf{b)} Recall confusion matrices for the network with a limited node degree of 20. Excess edges were uniformly sampled according to their length in the $wire \ vs \ time$ plane.} 
    \label{fig:max20}
\end{figure}

\begin{figure}
    \centering
    \includegraphics[scale=0.34]{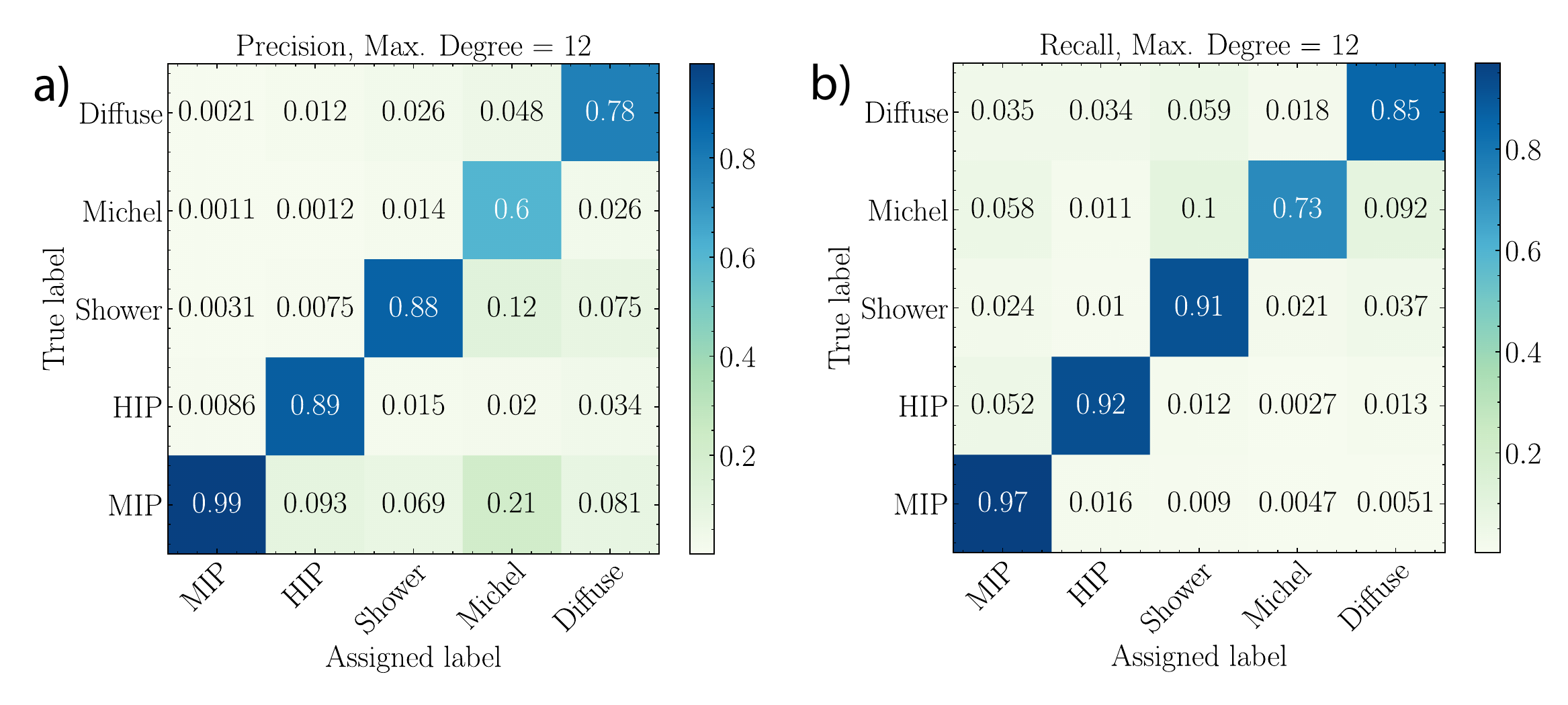}
    \caption{\textbf{a)} Precision and \textbf{b)} Recall confusion matrices for the network with a limited node degree of 12. Excess edges were uniformly sampled according to their length in the $wire \ vs \ time$ plane.} 
    \label{fig:max12}
\end{figure}

Figure \ref{fig:baseline} depicts the precision and recall matrices for the baseline Delaunay wiring. Precision is defined as $\mathrm{P = TP/(TP+FP)}$, while recall is defined as $\mathrm{R = TP/(TP+FN)}$. Figures \ref{fig:max20} and \ref{fig:max12} show these same matrices when the node degree is limited to 20 and 12, respectively. The excess edges are pruned by sorting the edges of each node whose degree is greater than 12 or 20 in the $wire \ vs \ time$ space according to their length, and uniformly sampling the resulting edge array. The uniform sampling ensures that edges of different length, representing distinct interaction scales, remain in the final wiring. 
There is no significant difference on the confusion matrices in the cases where the node degree is limited to 20 and the baseline wiring. However, when the node degree is limited to 12, there is a very small degradation of the network's performance. 

\begin{figure}
    \centering
    \includegraphics[scale=0.4, trim=0cm 2.5cm 0cm 1cm, clip]{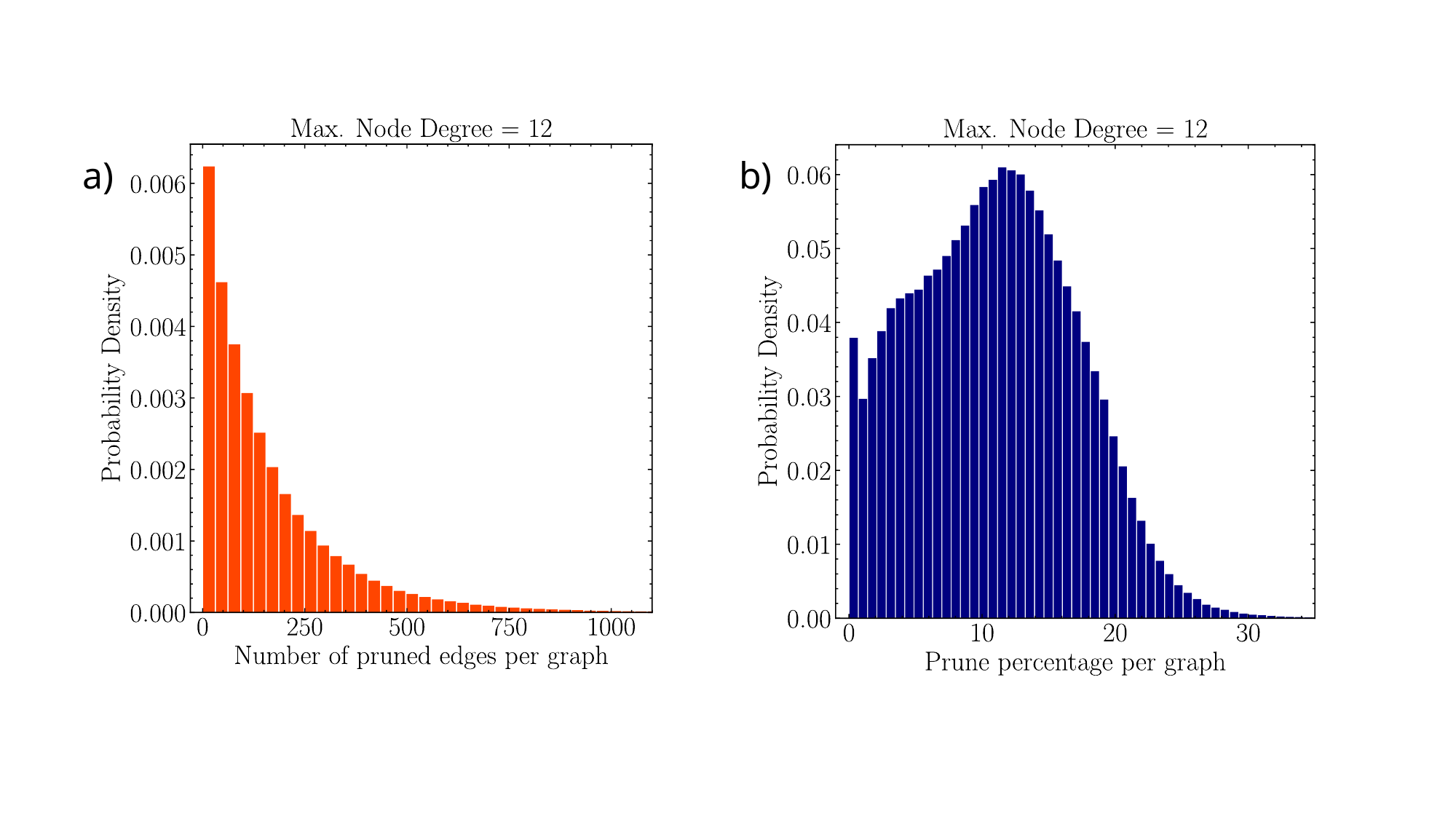}
    \caption{\textbf{a)} Total number of edges and \textbf{b)} percentage of edges pruned for the network with a limited node degree of 12. Excess edges were uniformly sampled according to their length in the $wire \ vs \ time$ plane. Plots are normalized to integrate to unity.}
    \label{fig:edgecount}
\end{figure}

Figure \ref{fig:edgecount}.a illustrates the number of pruned edges per graph while Figure \ref{fig:edgecount}.b presents the prune percentage per graph, when limiting the node degree to 12. In some graphs, the number of pruned edges can be as large as 800. We find that roughly 30\% of the nodes had a degree higher than 12, which indicates that some graphs in the dataset have a relatively dense edge matrix. On the other side of the spectrum, for a significant portion of the dataset the percentage of pruned edges was very low, smaller than 5\%, suggesting sparse edge matrices. However, for most graphs, about 13\% of their edges were removed. 

The percentage of pruned edges also carries some information about the underlying topology of the graphs present in the dataset. In most cases, the nodes with a very high degree are the ones in more ``isolated" regions in the wire $vs$ time space, with a low node density, as illustrated in Fig.~\ref{fig:example-event}, and they don't lie along HIP/MIP tracks. The unusually high degree of these hub nodes arises from their connection with all the nodes from a MIP or HIP track. Therefore, the percentage of pruned edges or, equivalently, the number of nodes with degree higher than $M$, stores high-level information about the graph's structure, and thus could be treated as a structural feature~\cite{physicsinjection}.

In summary, through this graph rewiring test, we quantified the level of redundancy present in the edge connections used in NuGraph2, supporting our hypothesis that this level of redundancy limits the applicability of mask-based explainability methods to NuGraph2. We also clarified the role of hub nodes, and evaluated the minimum number of connections they need to host to avoid degrading performance. These observation can also lead to improvements in the graph construction, that can potentially result in faster training and inference.

\subsection{Latent Space}
\label{sec:latentspace}

\begin{figure}
    \centering
    \includegraphics[width=0.9\linewidth]{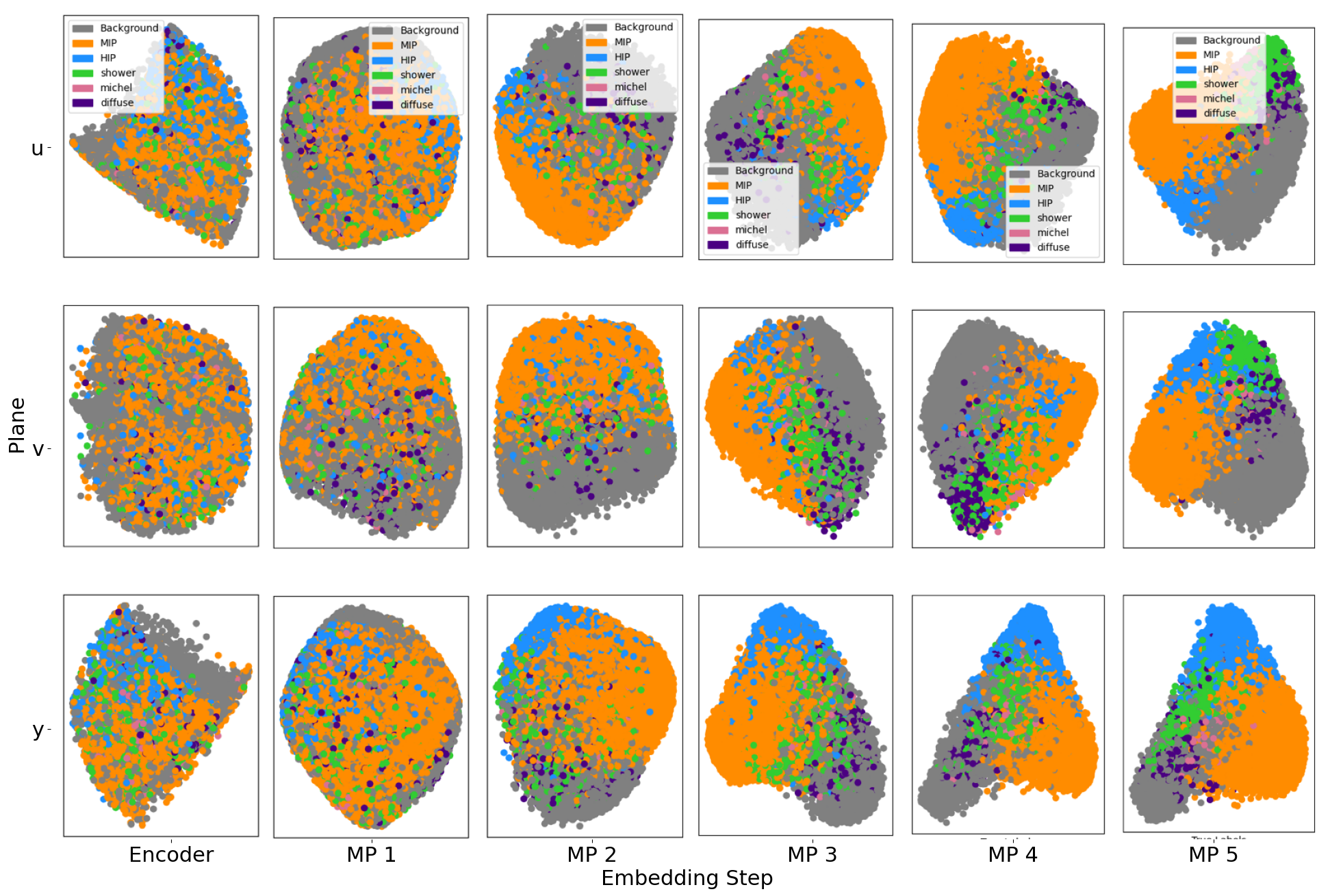}
    \caption{Embeddings of the network. After each major step in the network (encoder to message passing, between each nexus to planar message), the latent space is extracted and run through a 30 dimensional principal component transform. This plot shows the highest two dimensions as a function of progression through the network at each of the different places.}
    \label{fig:embeddings}
\end{figure}

To study the behavior of the network as a whole, we first examine the encoded space produced by different parts of the network and how they change over time. 
To do so, we take the fully trained network but stop inference after a given point in the network and view the output within an arbitrary downscaled space. 
The encoded space collapsed into a two-dimensional space is shown in Fig.~\ref{fig:embeddings} as a function of the network iteration. Here different colors are used based on the true semantic category of the nodes, so that we can see by eye at which step the network learns to separate between categories.
For a more quantitative analysis, we use a subsample of 25\% of the reserved validation data to produce a 30 dimensional PCA based decomposition which is used to calculate the separation of labels into "clusters" using a silhouette score.\cite{silhoutte_rousseeuw}
Silhouette scores are used based on their ability to show both cohesion and separation of a defined cluster, where a large value indicates a very well defined cluster. 
In this case,  we do not expect perfect separation of clusters due to the reduced dimensionality resulting from the PCA decomposition, but expect a change in score when clusters show better agreement.
The sample size and the number of dimensions were selected based on memory limitations of the used hardware. 
Resulting silhouette scores are reported in Table~\ref{tab:silhoutte}. 

Large change in the quality of clusters between steps allow us to see changes in the latent representation of labels and how well the latent space maps to labels - indicating when the encoding is stable.
The order of cluster formation for labels can similarly show us the successes and short-comings of the classification method. Clearly, the network identifies MIP and background nodes first, as they are the most represented categories in the dataset. On the other hand, it is not until the last two iterations that categories with smaller populations such as shower and diffuse stand out as a cluster. The fact that there are diminishing improvements from the fourth to the last iteration supports the choice of five total iterations. 

While a similar result can be achieved by simply performing a class-wise analysis using confusion matrices, this allows a more nuanced view of the problem showing the progression of the network's latent space at different steps of inference.
In summary, by studying the evolution of the latent space as the information flows through the GNN we are able to gain insight into the inference process and the ability of the network to perform its classification tasks. 

Notably, because of the 2D representation of the embeddings in figure \ref{fig:embeddings}, there is not perfect agreement between the visualization of the cluster and the silhouette scores.
The consistent downgrade in cluster clarity between the encoder stop and the first message passing step indicates that the immediate neighbors of a given node are not useful for classification, and the non-local nature of message passing is important to quality predictions. 
The large jump in cluster separation between 3 and 5 also supports the choice of 5 iterations as nominal configuration in NuGraph2~\cite{nugraph2}. Despite the relatively high number of message passing steps used, this is the optimal number of messages for this network.  
It does also show the failure of the network to encode Michel electrons, shown by the decrease in cluster clarity. 
Logically, this implies that the fewer number of Michel electrons are not weighted sufficiently within the dataset to overcome the contributions from other classes updating the node's hidden representation.

\begin{table}[]
    \centering
    \begin{tabular}{l | c | c }
        
        Label & Network Step & Mean Score \\
        \hline\hline

        HIP & Encoder & 0.0085 \\
        & 1 & 0.0052\\
         & 3 & 0.0065 \\
         & 5 & 0.0122 \\
        \hline
        MIP & Encoder & 0.0064 \\
        & 1 & 0.0056 \\
         & 3 & 0.0079 \\
         & 5 & 0.0113 \\
        \hline
        Shower & Encoder & 0.0078 \\
                & 1 & 0.006 \\
         & 3 & 0.0064 \\
         & 5 & 0.0103 \\
         \hline
        Michel & Encoder & 0.0098 \\
                & 1 & 0.0059 \\
         & 3 & 0.0055 \\
         & 5 & 0.0034 \\
        \hline
        Diffuse & Encoder & 0.0121 \\
                & 1 & 0.0047 \\
         & 3 & 0.006 \\
         & 5 & 0.0118 \\
        \hline
        Background & Encoder & 0.0088\\
        & 1 & 0.0049 \\
         & 3 & 0.008 \\
         & 5 & 0.0106 \\

    \end{tabular}
    \caption{
    Mean Silhouette score across planes for the clusters based on the particle meant to be included cluster. 
    Scores are calculated using a 30 dimensional decomposition and taking labels to be the class label of the embedding. 
    }
    \label{tab:silhoutte}
\end{table}

\subsection{Decoder Probes}
\label{sec:probes}

To assess the effectiveness of the network at different stages for different classification tasks, we establish different metrics across subsets of our dataset, which we refer to as concepts. 
While this method tests similar features to the examination of the latent space, and similarly accesses stages of inference, it is more flexible. 
Establishing different test metrics allows for much more flexibility in \textit{what} is being accessed. 
We are no longer constrained to just inherent labels in the dataset. 

These concepts represent emergent properties of the data that can be learned by the network. 
As an analogy, this can be thought of testing the ability of a convolutional network trained on images of zebras to recognize stripes as a generalized concept. 
This work is inspired by the notion of "Testing with Concept Activation Vectors" \cite{kim_interpretability_2018}. 
The key difference is that we do not use examples to construct a subset which represents our concept - rather, we produce a custom loss to measure how accurately selected concepts are learned. 
Using this method of decoder probes, this idea can be extended across other concepts that can be represented with a loss function. 
This means that this approach can be used to evaluate to what extent the network learns arbitrary concepts (which can be directly related to the primary learning task of the training, but not necessarily), thus effectively providing a tool to evaluate whether the network is gaining knowledge about underlying domain concepts. Such tool could be used e.g. to evaluate whether a given network shows properties similar to a foundation model.

To perform the study, we make single layer classification networks similar to the base NuGraph2's semantic decoder, and train these probes on the embedded node features. 
This probe training is independent of network training, and does not adjust the network's weights, meaning this is a purely post-hoc analysis method that can be applied to a given trained network (as long as the embedded features are accessible). 
Each network probe is a single layer fully connected network which takes the target layer's embedded features and output a five-dimensional vector representing class probabilities. 
Each concept is then expressed in terms of a specific loss function which describes how accurately the concept is captured in the latent space.

\begin{figure}
    \centering
    \includegraphics[width=0.6\linewidth]{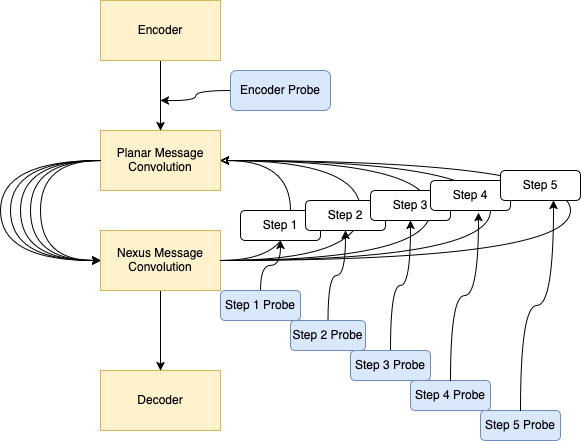}
    \caption{The placement of probes relative to the network architecture. Single dense layer "probes" are placed at different points in the network. Each one of these probes is trained after the network to evaluate how different combination of classes prefer as a function of different embedded spaces.}
    \label{fig:probe-design}
\end{figure}

Three test concepts were selected:  
two concepts represent relationships between specific types of particle hits within the graph, and one concept aims at showing how one semantic class is learned. 
All utilize a modification of cross entropy loss focusing on a certain semantic class behavior. 

\begin{itemize}
    \item Track-like particle detection. The two track-like particles (HIP and MIP) included in NuGraph2's classification are grouped into one class and tested for classification accuracy via the equation $$ L_M(\hat{y})= L(\hat{\bar{y}}, \bar{y}); 
    \bar{y}= 
        \begin{cases}
            1 & \text{if } y \in {\texttt{MIP}, \texttt{HIP}}\\
            0 & \text{if } y \not \in {\texttt{MIP}, \texttt{HIP}} 
        \end{cases}
    $$ 
        
    \item Track-like particle separation. HIP and MIP particles are isolated and accuracy is accessed on how well the two classes are separated, independent on how well the two are distinguished from the other classes, via the loss $$ L_M(\hat{y}) = L(\hat{y}_{y \in (\texttt{HIP}, \texttt{MIP})}, y_{y \in (\texttt{HIP}, \texttt{MIP})})$$
    
    \item Accuracy of only Michel Electrons (the least represented class). $$L_M(\hat{y}) = L(\hat{y}_{y=\texttt{Michel}}, y_{y=\texttt{Michel}}) $$
\end{itemize}

All of these concepts are easily represented by taking the loss of certain subsets, and does not require any specific extra labeling of data. However, the method is more general and can be applied to more arbitrary concepts, although they may require additional labels in the data set.

\begin{figure}
    \centering
            \includegraphics[width=.45\textwidth]{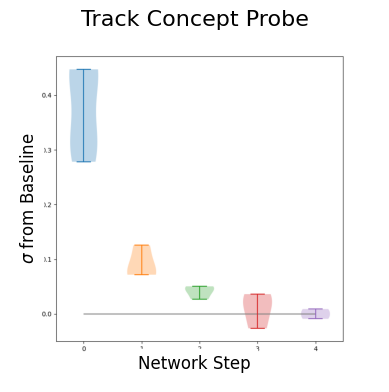}\hfill
            \includegraphics[width=.45\textwidth]{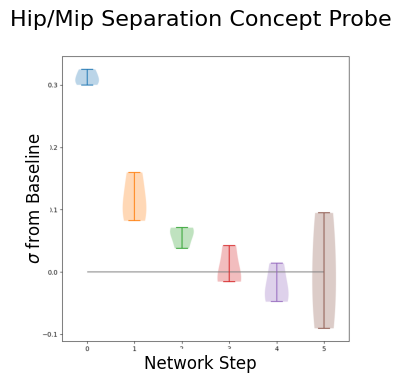}\\
            \includegraphics[width=.45\textwidth]{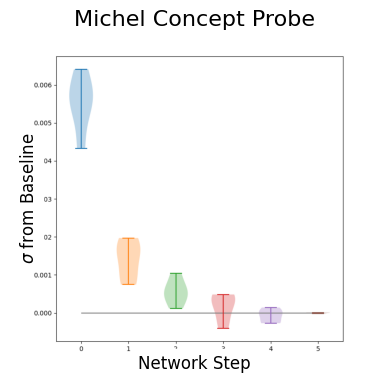}
    \caption{
        Concept Probes across the different stages of the network for 3 different concepts. 
        From left to right - Tracks Recognition, HIP/MIP Particle Separation, Michel Electron Identification. These probes are used to identify the amount of information about a "concept" is learned at each step, and as the network exhausts the data available the relative change from baseline drops. 
    }
    \label{fig:concept_probes}
\end{figure}

The probes are then trained in a k-fold schema, using a random subset of the original NuGraph2's training data.
The mean loss of the best performing probe is taken, and a loss plateau point is identified. 
This is used as a reference point against the other probes, see figure \ref{fig:concept_probes}. 
These plots can be used as a method for testing how quickly different concepts within the network are identified as a function of the message passing step of the network.
A sharp decrease in the validation loss in the early steps indicates the network has only a small amount of gains left to make in the later steps for this concept.

This method allows us to make the following inferences about NuGraph2's behavior: 

\begin{enumerate}
    \item Tracks are the "easiest" concept to identify among the concepts tested. This is judged by seeing the number of network steps needed before the probe losses plateau. We can see this in figure \ref{fig:concept_probes}, where losses are roughly stabilized by network step two. 
    This implies there is very little left to encode in the latent space regarding the relationship between HIP and MIP particles in the data the network has access to, and additional layers or operations applied to the data would not provide additional accuracy. 
    This can lead to the assumption that as this is the easiest concept to identify, it could also be the easiest concept to overlearn as well. 
    
    \item 
    The use of non-local information from the message passing steps helps both "difficult" and "easy" concepts, but not to the same degree. 
    There is no place in the network where we see a decrease in concept accuracy from the probes, nor a spot where the probes plateau only to grasp the concept afterwards.
    This implies there is not a number of hops where the nodes in the graph become actively dissimilar or unrelated for any of the tested concepts, decreasing the accuracy of the concept probes, nor is there a minimum number of hops needed to improve the concept accuracy.  
    While this is most likely not applicable to all problems, it is important to recognize that the construction of our problem results in a situation where non-local information, at least to 5 hops away, does not decrease the accuracy of these concepts.
        
    \item There are diminishing gains to be made for additional message passing steps, as can be argued from the fact that all of the probe losses follow a roughly exponential decay. 
    It is interesting to note that this feature is present both for concepts with high accuracy, such as classifying MIP and HIP, and concepts with relatively worse accuracy, such as Michel classification. The steps indicate the number of hops the information travels across the graph, so that early steps correspond to "local" information with respect to the node being evaluated, and in later steps less local information contributes. This study indicates that the relevant information needed by the network to make its prediction are mostly local in terms of node connectivity, and that distant information contributes meaningfully only if it is within a few hops.
    
\end{enumerate}

In summary, decoder probes are a novel tool to determine which concepts are represented in the network latent space and measure how quickly these concepts are identified during inference.
We present this method as a stepping stone for robust measurement of network learning outside of basic loss functions, presenting a more targeted approach to attributing informational flow.

\section{Conclusion}
\label{sec:conclusions}

As AI becomes more ubiquitous in scientific applications, we argue that interpretablity and explainability is more important than ever, yet underdeveloped in many types of networks. 
In testing on NuGraph2, we show the need for more methods that work on dynamic, heterogeneous networks. 
Despite testing standard methods (such as GNNExplainer), we found they are not a silver bullet for local explanations nor global information about the network. 
We also find these methods fail under the constraints of our network, shown by the work done examining the role node-degree has in prediction accuracy.
Therefore, both custom and generic methods are required to investigate the "how" and "why" of classification outputs. 

We show that while a network's expressivity can be a boon to reach higher accuracies, it can hinder efforts to trace a reasoning behind a result. Attempting to force a network to show connections that are "logical" to human understanding can be a fool's errand when this requirement is not directly enforced in the network itself. We investigated the information flow through the graph by performing a rewiring exercise that quantifies the level of redundancy in the network and clarifies the role of large multiplicity nodes, and showed that a densely connected network is not required for accurate predictions.
We further investigated how the latent space representation evolves as a function of the message passing iteration, demonstrating that separation of different classes in the latent space increases up the last iteration.  In our exploration of the network, we show how multiple methods are required to reach an understanding of an AI/ML's method's inner-workings. We introduced a method to assess whether specific concepts are captured in the latent space, allowing to test the learned space beyond ideas directly instilled by the loss function. This work advances explainability techniques towards addressing even more fundamental questions about the network behavior. 

\section{Acknowledgments}
The authors would like to thank Burt Holtzman and Maria Acosta for their management of computing resources used for running a large amount of the analysis included in this work. 
We acknowledge the MicroBooNE Collaboration for making publicly available the data sets~\cite{abratenko_2022_7261921, abratenko_2023_8370883} employed in this work. 

This work was produced by FermiForward Discovery Group, LLC under Contract No. 89243024CSC000002 with the U.S. Department of Energy, Office of Science, Office of High Energy Physics. Publisher acknowledges the U.S. Government license to provide public access under the DOE Public Access Plan DOE - \url{http://energy.gov/downloads/doe-public-access-plan}. This work was funded by a grant from the University of Illinois Discovery Partners Institute.
 \bibliographystyle{IEEEtran}
\bibliography{cite}

\end{document}

%% file: packages.tex
\usepackage[ascii]{inputenc}
\usepackage[T1]{fontenc}
\usepackage{geometry}

\usepackage{amsmath}
\usepackage{amssymb}
\usepackage[numbers,sort&compress]{natbib}
\usepackage[compact]{titlesec}
\usepackage{floatflt}
\usepackage{longtable}
\usepackage{wrapfig}
\usepackage{floatrow}
\usepackage{graphicx}
\usepackage{tabularx}
\usepackage{subcaption}
\usepackage{url}
\usepackage[usenames,dvipsnames]{color}
\usepackage{pdfpages}
\usepackage{enumitem}
\usepackage{authblk}

\usepackage{mathptmx}
\usepackage[english]{babel}
\usepackage{amsmath}
\usepackage{amssymb,amsfonts,textcomp}
\usepackage{array}
\usepackage{supertabular}
\usepackage{multirow}
\usepackage{hhline}
\usepackage{enumitem}
\usepackage{xspace}
\usepackage{upgreek}
\usepackage{microtype}
\usepackage{lineno}
\usepackage{stackrel}
\usepackage{accents}

\makeatletter
\newcommand\arraybslash{\let\\\@arraycr}
\makeatother

%% file: authors.tex
\author{%
    M. Voetberg$^1$\footnote{\texttt{maggiev@fnal.gov}}, 
    Vitor F. Grizzi$^{2, 4}$\footnote{Now at Argonne National Lab}, 
    Giuseppe Cerati$^1$, 
    Hadi Meidani$^3$, 
    V Hewes$^5$

    \small{
    $^1$ Department of Nuclear, Plasma, and Radiological Engineering, University of Illinois Urbana-Champaign, Urbana, IL 61801
    $^2$ Chemical Sciences and Engineering Division, Argonne National Laboratory, Lemont, IL 60439
    $^3$Fermi National Accelerator Laboratory,  Batavia, IL 60510, USA
    $^4$ Department of Civil and Environmental Engineering,  University of Illinois Urbana-Champaign, Urbana, IL 61801
    $^5$ Department of Physics,  University of Cincinnati, Cincinnati, OH 45221-0377
}
}